\documentclass[aps,rmp,amsfonts,amssymb,nofootinbib,floats,longbibliography,twocolumn, superscriptaddress]{revtex4-2}
\usepackage{graphics}
\usepackage{epsfig}
\usepackage{xcolor}
\usepackage{mathtools}
\usepackage{mathrsfs,bm}
\usepackage{changes}
\usepackage{soul}

\definecolor{dodgerblue}{HTML}{1E90FF}
\definecolor{columbiablue}{HTML}{87AFC7}
 
\usepackage[colorlinks=true,citecolor=dodgerblue,linkcolor=dodgerblue,urlcolor=dodgerblue]{hyperref}

\usepackage{booktabs}

\setcitestyle{square,comma,numbers}
\makeatletter
\def\NAT@cmprs{\@ne}
\makeatother

\let \oldbm \bm
\renewcommand{\vec}[1]{\oldbm{#1}}

\newcommand{\br}{{\bf r}}
\newcommand{\bk}{{\bf k}}

\newcommand{\proj}[1]{\left| #1 \right> \left< #1 \right|}

\newcommand{\ev}[1]{\langle #1 \rangle}
\newcommand{\ket}[1]{\left| #1 \right>}

\renewcommand{\Im}{{\rm Im~}}
\renewcommand{\Re}{{\rm Re~}}

\def\bk{{\vec k}}

\def\bv{{\vec v}}

\def\bm{{\vec m}}

\def\br{{\vec r}}

\def\tr{\mathop{\mathrm{tr}}}

\def\Q{\mathcal{Q}}

\def\E{\mathcal{E}}

\def\V{\mathcal{V}}

\def\inv{^{-1}}

\def\A{\mathcal{A}}

\begin{abstract}
Electronic properties of quantum materials solids are often well understood via the low energy dispersion of Bloch bands, motivating single band approximations in many metals and semiconductors. However, a closer look reveals length and time scales introduced by quantum dipole fluctuations due to interband mixing, which are reflected in the momentum space textures of the electronic wavefunctions. This structure is usually referred to as quantum geometry. These new scales not only qualitatively modify the linear and nonlinear responses of a material but can also have a vital role in determining the many-body ground state at low temperatures. In this Perspective, we explore how quantum geometry impacts properties of materials and outline recent experimental advances that have begun to explore quantum geometric effects in various condensed matter platforms. We discuss the separation of scales that can allow us to estimate the significance of quantum geometry in various response functions. 
\end{abstract}

\begin{document}
\title{Quantum Geometry and the Hidden Scales in Materials}

\author{Nishchhal \surname{Verma} }
\affiliation{\mbox{Department of Physics, Columbia University, New York, NY 10027, USA}}
\author{Philip J. W. Moll}
\affiliation{\mbox{Max Planck Institute for the Structure and Dynamics of Matter, Hamburg 22761, Germany}}
\author{Tobias Holder}
\affiliation{\mbox{School of Physics and Astronomy, Tel Aviv University, Tel Aviv, Israel}}
\author{Raquel \surname{Queiroz} }
\affiliation{\mbox{Department of Physics, Columbia University, New York, NY 10027, USA. }}
\email{raquel.queiroz@columbia.edu}

\maketitle
\section*{Introduction}
Understanding the electronic properties of quantum materials relies on determining the relevant energy, length, and time scales to build a suitable model. In a crystalline solid, the fundamental length scale is the atomic lattice constant $a$, which allows for the definition of crystal momentum $k$ that labels both band energies $\varepsilon(k)$ and wavefunctions ${\psi(k)}$. Much of the success in describing metals and semiconductors lies in the fact that electron transport can be described semiclassically. A phenomenological length scale, the electronic mean free path, captures the long wavelength behaviour. Because this mean-free path is typically much larger than the lattice constant, the structure of materials at the lattice scale $a$ is considered unimportant to the collective behavior at low-energies, justifying long-wavelength approximations and hydrodynamic descriptions. On the other end of the scale, attempting to correctly capture the effect of strong local interactions, microscopically motivated lattice models, like the Hubbard model,  are often simplified by taking a single site per unit cell. In both of these regimes, a complicated quantum material problem is reduced to a single-band approximation, ignoring quantum interference effects across distinct orbits of the (often intricate) material unit cell. 

However, orbital mixing can lead to surprising qualitative changes in the physical behavior of materials.The mixing within and across unit cells is the source of new length scales that come not from how band energies change, but by how electron wavefunctions change with crystal momentum ${\partial_k\psi(k)}$. These changes are broadly referred to as the quantum geometry of solids. Apart from changes in the wavefunctions, a very convenient way to interpret quantum geometry in solids is as ground state dipole fluctuations \cite{Resta2011}, generalizing to infinitely large systems the well-understood case of dipole fluctuations in isolated atoms due to virtual level transitions \cite{Traini.Traini.1996}. Generally, these fluctuations can be used to characterize the size, shape, and angular momentum of atomic orbitals. Strikingly, on the lattice, qualitatively new orbitals can emerge from the quantum interference of distinct neighboring sites and are substantially richer than the isolated atomic problem. Dipole fluctuations have dramatic effects in linear and nonlinear responses of common materials, particularly those in low spatial dimensions such as graphene, transition metal dichalcogenides, or moiré heterostructures.

Over the last few years, with the rapid advances in the field of two-dimensional heterostructures, there has been mounting evidence that quantum materials can exhibit distinct transport properties, such as superfluid or exciton stiffness, solely due to differences in their quantum geometry \cite{rossi2021quantum,Yu2024QuantumGeometry}. This effect is widespread and can be appreciated by a simple example: rock salt and diamond reflect light very differently. Their energy gaps are comparable, but their electronic wavefunctions are remarkably different. The effects of wavefunction geometry on the response of materials can be particularly striking in flat-band systems, where they can qualitatively change the physical properties. Such flat bands can now be achieved by moiré engineering in a variety of materials \cite{JulkuPRB2020, hu2019, XiePRL2020}. Therefore, identifying observables capable of directly measuring or quantifying quantum geometry in materials has become an important goal in the field. Quantum geometry can also play a fundamental role in the competition between correlated states, where the quantum geometry of the normal state can help stabilize exotic quantum orders such as fractionalized topological phases and superconductivity.

In this Perspective, it is our goal to emphasize that quantum geometry is an essential feature of any model with multiple bands. We aim to highlight that quantum geometry can be used in solids to quantitatively and systematically describe the response of bound electrons in infinite lattices. As we argue here, this framework is valid even for metals where only part of the electric charge is bound, coexisting with itinerant plane-wave electron states. Furthermore, the quantum geometric description of bound electrons is straightforwardly applied to systems without translation symmetry and beyond the single particle approximation. Therefore, it is a universal feature that systematically tracks the length and time scales that capture dipole dynamics in solids. In the following, we explain the concept of band geometry, focusing on its real space interpretation, and its immediate consequences for the physical properties of quantum matter. We further discuss possible experimental probes to measure the geometric tensor and highlight some of the intrinsic difficulties in their measurement.

\begin{figure}
    \centering
   \includegraphics[width = \columnwidth]{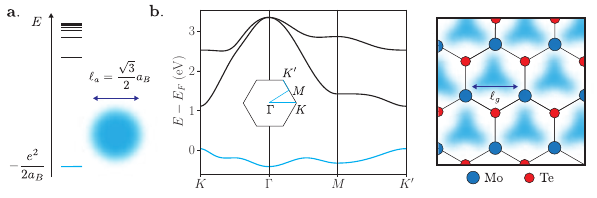}
    \caption{
    Dipole fluctuations lead to uncertainty in electron position. The corresponding spatial spread is characterized by the length scale $\ell_g$ determined by the quantum geometry of the ground state.
    {\bf a}. The hydrogen atom, where the ground state wavefunction is an $s$ orbital with spread $\ell_a^2= 3 a_B^2/4$ where $a_B$ is the Bohr radius. {\bf b}. Monolayer ${\rm MoTe}_2$, a semiconductor with $\approx 1\rm eV$ band gap and a topological obtruction. The localized electrons are shared between neighboring sites with a large spread spanning multiple unit cells due to nontrivial orbital interference. The geometric scale $\ell_g$ is comparable to the lattice constant.
    }
    \label{fig:dipole transitions}
\end{figure}

  \section*{What is Quantum Geometry?}

  When a quantum state adiabatically moves through a parameter space to make a closed loop, interference effects can cause the state to pick up a phase difference, called the geometric phase. This can take many forms, from the polarization of light \cite{Pancharatnam1956}, threading of a magnetic flux \cite{Bohm.Aharonov.1959}, or changes in atomic positions of molecules \cite{Longuet-Higgins.Herzberg.1963}. Geometric phases are understood to be general in quantum systems \cite{berry1984}, arising from the Riemannian structure of the projected manifold of occupied quantum states. The quantum geometric tensor (QGT)\cite{Vallee.Provost.1980} can be used to describe the adiabatic evolution in parameter space, showing that it is characterized not only by a phase rotation but also by a loss of the projected norm of the quantum states. Both effects are simultaneously characterized by the real and imaginary parts of the QGT. The imaginary part of the QGT, which describes the Berry curvature, or Berry phase, is well understood in electronic structure theory \cite{Xiao2010}.  However, the effects of the real part, the quantum metric, has remained mostly unexplored. Nonetheless, just as its imaginary counterpart, the real part is bound to play a significant role in quantum states of matter.

\subsection*{Fidelity susceptibility} 

The definition of the QGT is universal and has only two requisites: a parameter space $\{\lambda_\mu\}$ and a projector $\hat{P}$, defining a subset of quantum states in the Hilbert space,
\begin{equation}
    \hat{\mathcal{Q}}_{\mu\nu} =
    \hat{P}\partial_\mu\hat{P} \partial_\nu\hat{P}\label{eq:qgt}
\end{equation}
where $\partial_\mu \equiv \partial/\partial_{\lambda_\mu}$. We use the symbol without a hat $\Q_{\mu\nu}\!=\!\tr[\hat\Q_{\mu\nu}]$ to indicate the trace over the internal states of $\hat P$.
Consider the projector into a single state of an Hamiltonian $\mathcal{H}$, $\hat{P} = |\psi \rangle \langle \psi|$. When perturbations $\mathcal{H}+\sum_\mu\lambda_\mu \mathcal{H}_\mu'$ are introduced, the parameters $\lambda_\mu$ define a parameter space and Eq.~\eqref{eq:qgt} its geometric tensor. In this case, the QGT captures the sensitivity of the  state $\ket\psi$ to perturbations, or fidelity susceptibility \cite{kolodrubetz2017geometry}. 
The fundamental reason behind quantum geometry is that the wavefunction overlap at adjacent points in parameter space deviates from unity, even when the wavefunctions are normalized at each point.
Michael Berry \cite{berry1984} demonstrated that wavefunctions acquire infinitesimal phases during this evolution, $\langle \psi(d\vec\lambda) | \psi \rangle \approx 1 - i d\lambda_\mu \langle \psi | \partial_\mu \psi\rangle$, that accumulate to a gauge-invariant geometric phase when the path is closed. The phase is captured by the Berry curvature $\Omega_{\mu\nu} = i\epsilon_{\mu\nu}\partial_\mu \langle \psi | \partial_\nu \psi\rangle$. 
Concurrently, the reduction in magnitude of the wavefunction overlap is also gauge-invariant \cite{Fubini.Fubini.1904,Study.Study.1905}. Expanding the wavefunction overlap to second order, we find 
\begin{equation}
|\langle \psi(d\vec\lambda) | \psi \rangle|^2 = 1 - g_{\mu\nu} d\lambda_\mu d\lambda_\nu + \mathcal{O}(d\lambda^4),
\end{equation}
where $g_{\mu\nu}$ is the quantum metric that quantifies the distance between nearby quantum states.
The QGT tensor signals where wavefunctions change abruptly, such as during quantum phase transitions \cite{Vojta.Vojta.2003}, conical intersections of energy landscapes \cite{Longuet-Higgins.Herzberg.1963} or near singular band crossings in electronic band structure \cite{Vishwanath.Armitage.2018}, at which points the susceptibility (Eq.~\eqref{eq:qgt}) diverges \cite{Gu2008, rattacaso2020quantum}.

An elegant application of the geometric tensor for quantum materials was introduced by Walter Kohn in the 1960s \cite{Kohn.Kohn.1964}, who studied a many-body ground state $\hat P^0=\proj{\psi_0}$ under twisted boundary conditions. The twist shifts the momentum in the spatial direction $\mu$ by the adiabatic parameter $\kappa_\mu$, $p_\mu \rightarrow p_\mu - \hbar \kappa_\mu$. Kohn argued that in the thermodynamic limit, the susceptibility, as given by the QGT (Eq.~\eqref{eq:qgt}) distinguishes the ground state of a metal from an insulator, diverging in metals while remaining finite for insulators \cite{resta2017geometrical}. 
This distinction based on the localization of wavefunctions bypasses the reference to a spectral gap, which can be ill-defined in disordered and many-body systems.

\subsection*{Dipole fluctuations} 

The momentum shift under twisted boundary conditions can be interpreted as a susceptibility towards dipole transitions between states in $\hat P$ and states in the complementary projector $1-\hat P$. Since position and momentum are conjugate variables, $\partial_\mu \hat{P}\equiv i[\hat{r}_\mu, \hat{P}]$, we may express Eq.~\eqref{eq:qgt} as a localization tensor\cite{Sorella.Resta.1999, Martin.Souza.1999}:
\begin{equation}
    \mathcal{Q}_{\mu\nu} = {\rm tr} \left[\hat{P} (i[\hat{r}_\mu, \hat{P}]) (i[\hat{r}_\nu, \hat{P}]) \right] = \ev{\hat{r}_\mu(1-\hat P) \hat{r}_\nu}\label{eq:def-Q-r-general}
\end{equation}
where we have defined $\ev{\cdot}\!\equiv\!\tr[\hat P\cdot]$ as the expectation value at $T=0$.
As the localization tensor contains the dipole transitions between the ground state and excited states, an insulator will have nonzero quantum geometry as long as the ground state is fully localized, and not a position eigenstate.
The quantum metric, $g_{\mu\nu}=\Re\Q_{\mu\nu}$, introduces a length scale $\ell
_g = \sqrt{\tr g}$ associated with the size of dipole fluctuations. 

A straightforward yet instructive example to view the quantum metric as a measure of dipole fluctuations is to consider an isolated hydrogen atom subjected to an external electric field. The perturbed Hamiltonian, $H^\prime = H - e \hat{x} E$ (for a field along $\hat{x}$) modifies the ground state wavefunction from $|\psi_0\rangle$ to $|\psi_0(E)\rangle$. 
The quantum metric is related to the fidelity via $g_{xx} = \partial |\langle \psi_0(E) | \psi_0 \rangle|^2/\partial E^2$ as $E \rightarrow 0$. Using second-order perturbation theory, the metric becomes $g_{xx} = \sum_{m\neq 0}|\langle \psi_0 | \hat{x} |\psi_m \rangle|^2 = \langle (\hat{x} - \langle \hat{x}\rangle)^2 \rangle_0$. Therefore the metric measures the dipole fluctuations, which for the ground state of the hydrogen atom corresponds to the spatial spread of the $s$ atomic orbital, $\ell_g^2\!=\!(3/4)a_B^2$, with $a_B$ the Bohr radius. This example shows us that in a single atom, the size of the atomic orbital is a scale introduced by the geometric tensor.

The localization tensor (Eq.~\eqref{eq:def-Q-r-general}) is useful in that it can be applied to macroscopic systems where the behavior of electrons is qualitatively different from those in the hydrogen atom. Quantum geometry provides us with a language to capture the scale $\ell_g$ of dipole fluctuations of bound electrons in materials in infinite lattices, whose wavefunctions in the clean limit are fully delocalized Bloch states (Fig.\ref{fig:dipole transitions}).
The geometric length scale $ \ell_g $ is typically on the order of the lattice constant ($ \ell_g \approx a $). However, there are important exceptions where $\ell_g \gg a$. In a massive Dirac system, for instance, $\ell_g$ diverges as the energy gap closes and can remain much larger than the lattice constant for a small mass. This behaviour  underscores that $\ell_g$ arises from wavefunction interference across many sites and is not solely determined by local energetics.

In topological systems, such as those exhibiting quantized Hall conductivity~\cite{Pepper.Klitzing.1980,Nijs.Thouless.1982} or symmetry-protected boundary modes~\cite{Kane.Hasan.2010}, topological invariants place a generic lower bound on the magnitude of dipole fluctuations. 
This bound becomes saturated in idealized scenarios, such as free electrons in a magnetic field where the geometric scale coincides with the magnetic length $\ell_g^2 = \ell_B^2 = \hbar / eB$ \cite{Roy2014}. In trivial systems with weakly hybridized orbitals, the geometric scale captures the sizes of the orbitals themselves $\ell_g\approx a_B$, and can often be ignored, as is the case in single band approximations.
Crucially, the scale $\ell_g$ is universally a well-defined indicator of dipole fluctuations.

\begin{figure}
\centering
\includegraphics[width=\columnwidth]{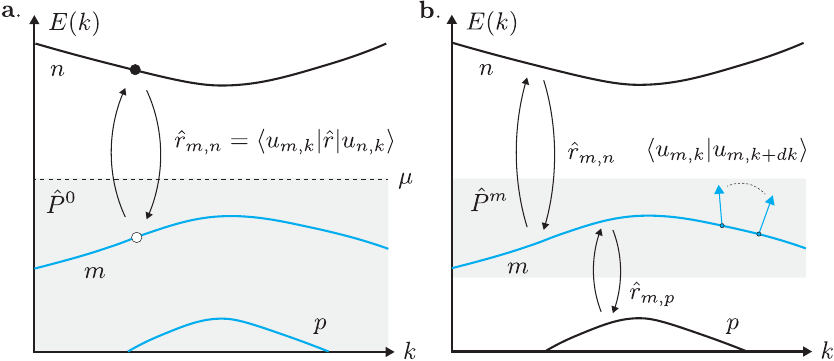}
\caption{
Choice of projected manifold in defining quantum metric. {\bf a}. The Kohn metric is defined using the ground-state projector $ \hat{P}^0 $ and captures dipole transitions between occupied and unoccupied states. {\bf b}. The quantum metric for a single isolated band $ m $ is defined using the band-resolved projector $ \hat{P}^m $, incorporating dipole transitions between band $ m $ and all other bands $ n \neq m $. The infinitesimal overlap between Bloch states is given by $ |\langle u_{m, \boldsymbol{k}} | u_{m, \boldsymbol{k} + d\boldsymbol{k}} \rangle|^2 = 1 - g^m_{\mu\nu} \, dk_\mu \, dk_\nu $, where $ g^m_{\mu\nu} = \sum_{n \neq m} \langle u_{m, \boldsymbol{k}} | \hat{r}_\mu | u_{n, \boldsymbol{k}} \rangle \langle u_{n, \boldsymbol{k}} | \hat{r}_\nu | u_{m, \boldsymbol{k}} \rangle $. The Kohn metric appears in optical sum rules, while the single-band metric governs the superfluid stiffness in flat-band systems. The two are formally distinct, in that they differ in their choice of the projected manifold, and are thus  not interchangeable.
}
\label{fig:matrixelements}
\end{figure}

\section*{Quantum Geometry in Band Theory}

The localization tensor (Eq.~\eqref{eq:def-Q-r-general}), quantifies the dipole matrix elements between the ground state and the excited states.
Even in the case of non-interacting electrons, this formalism is often impractical because of the vast Hilbert space.
However, from the simple case of an isolated hydrogen atom, we can gain valuable intuition about the effects of dipole fluctuations and derive a geometric framework to apply in systems with multiple electrons. This approach requires a careful definition of the dipole operator in an infinite periodic system, which is provided by the modern theory of polarization \cite{KingSmith1993}.
Due to Bloch's theorem, the electronic wavefunction in a periodic lattice potential can be decomposed into a plane-wave and a cell-periodic part, $|\psi_{n\bk}\rangle = e^{i \bk \cdot \hat{\br} } | u_{n\bk}\rangle$, where $n$ is the band index, $\bk$  the crystal momentum and $\hat\br$ the position operator. Being eigenstates of a Hermitian operator, the Bloch states are orthogonal. However, this orthogonality does extend to the cell-periodic function $|u_{n\bk}\rangle$ which can have nonzero overlaps for different momenta. This introduces a notion of distance between two cell periodic states, giving rise to a Riemannian structure within the Brillouin Zone (BZ) by the adiabatic parameter $\bk$ \cite{Zak.Zak.1989}.  

To be more precise, we note that the quantum geometric tensor is the dipole correlation function that requires a projector. A natural choice is the projector to a Bloch state resolved in both band index and crystal momentum $\hat P_{\bk}^n$. However, it is important to notice that the dipole matrix element 
    $\hat{r}_\mu^{mn}(\bk,\bk') = \langle u_{m\bk} | \hat{r}_\mu | u_{n\bk'} \rangle$
contains both intra- and interband contributions~\cite{Blount1962} (Fig.~\ref{fig:matrixelements}),  
\begin{equation}
    \hat{r}_\mu^{mn}(\bk,\bk')  = - \delta_{\bk,\bk'} \left[\A_\mu(\bk)\right]_{mn}+ i\delta_{mn}\partial_\mu \delta_{\bk,\bk'} \ ,
    \label{eq:blount-matrix-element}
\end{equation}
containing the matrix elements of the Berry connection $\left[\A_\mu(\bk)\right]_{mn}=i\langle u_{m\bk} | \partial_\mu  u_{n\bk} \rangle$, which are not by themselves gauge invariant. A crucial aspect of the quantum geometric tensor is that the complementary projector in Eq.~\eqref{eq:def-Q-r-general} removes the diagonal contributions of the position operator, and therefore, it is fully gauge independent and measurable. For Bloch electrons, the QGT can be compactly written in the band and momentum basis as
\begin{align}
    \Q_{\mu\nu}^{n}(\bk)=\left\langle\partial_{\mu} u_{n\bk}\right| 1-\hat P_{\bk}^n \left|\partial_{\nu} u_{n\bk}\right\rangle\label{eq:band_qgt}.
\end{align}
From $\Q_{\mu\nu}^n(\bk)$ one can similarly define a momentum resolved quantum metric $g^n_{\mu\nu}(\bk)=\Re[\Q^n_{\mu\nu}(\bk)]$ and Berry curvature $\Omega^n_{\mu\nu}(\bk)=2\Im[\Q^n_{\mu\nu}(\bk)]$ in the BZ. 
These two control the overlap and geometric phase between two infinitesimally close cell-periodic states $|u_{n,\bk}\rangle$ and $|u_{n,\bk+ d\bk}\rangle$ respectively.

Berry curvature effects are well known in the semiclassical theory of transport in terms of the anomalous velocity \cite{Xiao2010}. Originally derived through second-order perturbation theory \cite{Karplus1954}, the anomalous velocity was later recognized as a consequence of wavefunction overlaps. By explicitly deriving the equations of motion for a wave packet in the presence of an electric field, it has been demonstrated that the intrinsic anomalous velocity is precisely the Berry curvature \cite{Chang1996, Sundaram1999}.
Understanding the connection between the anomalous velocity and wavefunction overlap has revolutionized our understanding of the Anomalous Hall effect, offering a fresh perspective distinct from conventional mechanisms based on scattering theory \cite{Ong.Nagaosa.2010}. Furthermore, it has opened pathways to quantized Hall conductance by utilizing the quantization of the integral of Berry curvature over one band. This stems from the relation $\int_{\rm BZ} {\rm Im}[\Q^n_{\mu\nu}(\bk)] = 4\pi \mathcal{C}\epsilon_{\mu\nu}$ where $\int_{\rm BZ}\equiv\int d^d\bk/(2\pi)^d$ and $\mathcal{C}$ is the Chern number that can only be an integer.

In contrast, the quantum metric has remained largely unexplored, primarily due to the lack of an intuitive understanding. Its effects on semiclassical equations of motions are only now being explored \cite{Park2025quantumgeometry}. Moreover, challenges inherent in its definition create additional conceptual hurdles that complicate its interpretation and practical application.
The band resolved QGT (Eq.~\eqref{eq:band_qgt}) corresponds to dipole transitions between the band $n$ and all other bands $m \neq n$, filled or empty, at a given point in the BZ. It is thus qualitatively different from the metric introduced by Kohn, where the projector in Eq.~\eqref{eq:def-Q-r-general} is the ground state projector $\hat P^0$ (Fig.\ref{fig:matrixelements}). For band electrons, we can define $\hat P^0$ into all occupied single-particle states as
\begin{equation}\textstyle
    \hat{P}^0 = \int_{\rm BZ} \sum_n \Theta(\mu - \varepsilon_{n\bk}) | u_{n\bk}\rangle \langle u_{n\bk}|\label{eq:band_groundstateproj}
\end{equation}
with $\mu$ the chemical potential. In particular, in the case of multiple filled bands, the ground state geometric tensor cannot be obtained by summing $\Q^n_{\mu\nu}(\bk)$ over filled bands and momenta. This is because, unlike Berry curvature, the quantum metric is not band additive \cite{Ozawa2021}.

\begin{figure}
    \centering
    \includegraphics[width=\linewidth]{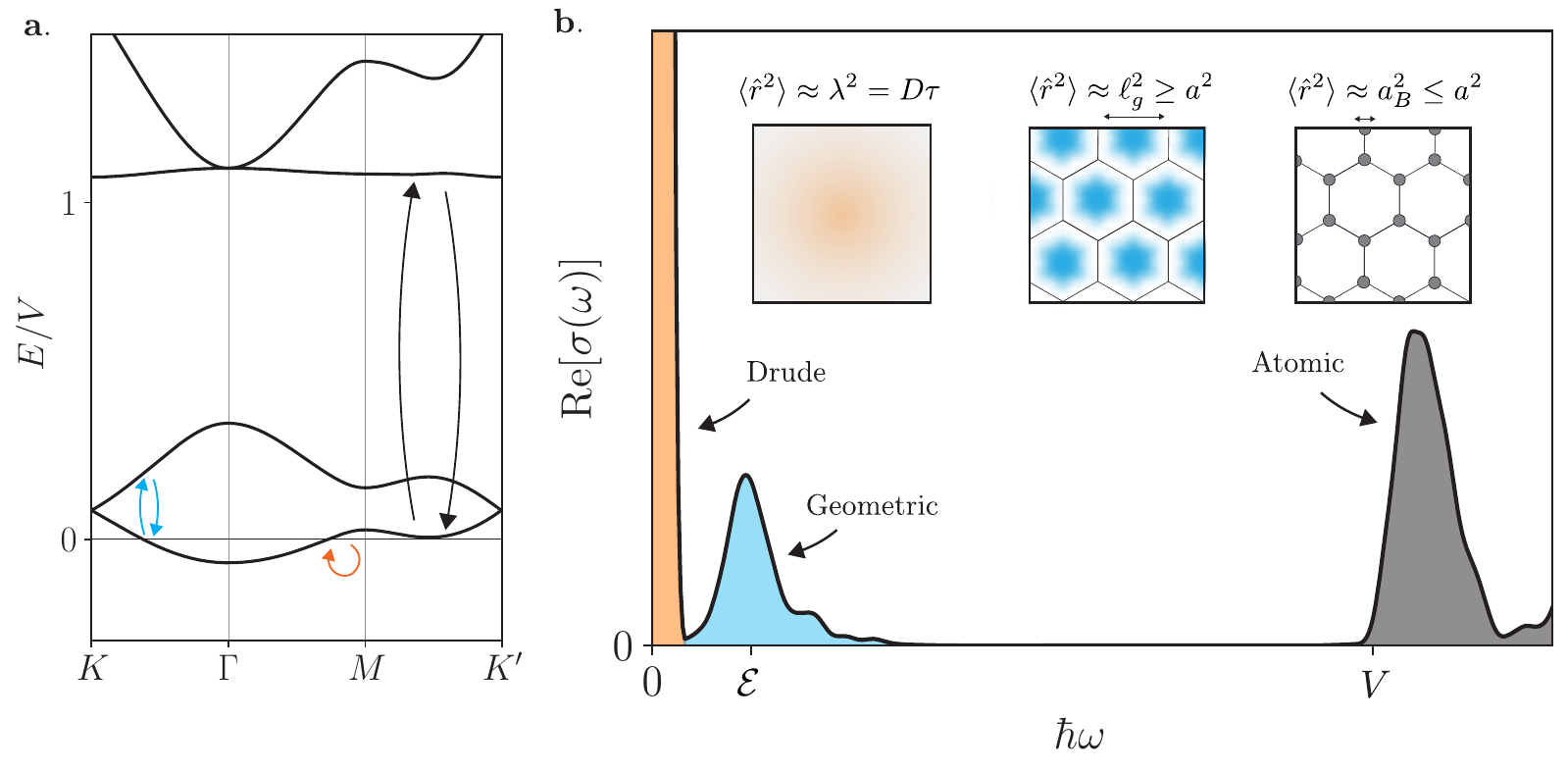}
    \caption{Separation of scales and the emergence of a geometric length scale from lattice interference. {\bf a}. Band structure for a free particle in a periodic honeycomb lattice potential. The tight-binding approximation is valid when the potential $V$ is deep enough to spectrally isolate a few orbitals, here the two lower connected bands. The arrows indicate optical transitions within a single band (orange), within the low-energy band manifold (blue), and between the low- and high-energy bands (black). {\bf b}. Optical conductivity for the same system, where low-frequency spectral weight features a Drude weight due to the Fermi surface (orange) in addition to a geometric spectral weight (blue), due to optical transitions between the same orbital across distinct lattice sites. The higher-frequency features reflect dipole transitions that correspond to fluctuations at the scale of an atomic orbital, truncated away in the tight-binding approximation. Looking at the spectral weight at low energies (blue) we can define an effective geometric length scale, $\ell_g$, that governs dipole fluctuations due to lattice interference, as well as a resonant energy $\E$ associated with its local dynamics. The resonant energy $\E$ is defined by a weighted average of band energy differences $\varepsilon_{n\bk}-\varepsilon_{m\bk}$ by geometric factors and can often be approximated by the first dominant peak in optical conductivity, comparable to the hopping energy. The boxes indicate the characteristic length scales associated with the dipole fluctuations contributing to the long-wavelength (orange), intermediate (blue), and atomic-scale (black) spectral weight.
    }
    \label{fig:separation-scales}
\end{figure}

The fact that the quantum metric is not additive can be explained via Wannier functions. The integral of the quantum metric over the Brillouin zone (BZ) for a single band, $\int_{\rm BZ} {\rm Re}[\mathcal{Q}^n_{\mu\nu}(\bk)] = g^n_{\mu\nu}$, sets a lower bound on the spatial spread of the Wannier function \cite{Vanderbilt.Marzari.1997, MarzariRMP2012}. 
When additional bands are included in the construction of the Wannier function, it is reasonable to expect a more localized Wannier state with a reduced spatial spread. However, since the quantum metric is inherently a positive quantity, the sum of the metric of individual bands within a projected manifold cannot be less than the metric of the full manifold. Thus the metric cannot be additive, $g_{\mu\nu} \neq \sum_{n \in {\rm occ.} } g^n_{\mu\nu}$. The metric is consequently not a conventional response function and depends on both filled and unfilled bands \cite{verma2024step}. It is important to distinguish this measure of spread from exponential localization; for instance, in topological bands, Wannier functions are obstructed and exhibit power-law tails, even though their spatial spread as measured by the quantum metric remains finite. Therefore, also for topological bands the geometric scales introduced by dipole fluctuations are finite and well defined.

Setting these subtleties aside, the quantum metric and Berry curvature have a close relationship that becomes evident in the case of Landau levels.
As we know, the Chern number arises from the non-commutativity of the projected position operators in the lowest Landau level, $\hat{X} = \hat{P} \hat{x} \hat{P}$ and $\hat{Y} = \hat{P} \hat{y} \hat{P}$, with $\ev{[\hat{X}, \hat{Y}]} = i \ell_B^2$, where $\ell_B$ is the magnetic length. The non-commutativity of these operators naturally implies that the variance $\langle \hat{X}^2 + \hat{Y}^2 \rangle$ is subject to a lower bound. Remarkably, the lowest Landau level satisfies the so-called trace condition~\cite{Roy2014}, in which this variance saturates the bound.
In simpler terms, the cyclotron orbits of an electron in a Landau level have the smallest spatial spread capable of supporting the necessary winding.

Although real materials deviate significantly from the idealized limit of Landau levels, the connection between topology and geometry remains true. Topological bands inherently require extended Wannier states, as the quantum metric is subject to a lower bound imposed by the topological index. Thus topology is a sufficient condition for non-trivial quantum geometry. This statement can be made more precise through an estimate of the quantum metric derived from the dielectric constant. Within the space of all insulators, it has been shown that the quantum metric in topological insulators is significantly larger than that in trivial insulators with comparable band gaps \cite{komissarov2024quantum}.
Importantly, topology is sufficient but not necessary for quantum geometry in real materials. Unlike the Berry curvature, the quantum metric is not restricted to systems that break time-reversal or inversion symmetry. Quantum geometry is present in all known materials in various amounts, and the key challenge is to quantify its influence on physical observables (Box 1). 

\section*{Where to find the quantum metric?}

The experimental challenge in measuring the metric is two-fold. First, the probe must couple to the symmetric part of the QGT, and second, it must identify the specific geometric scale being probed. This issue ties back to the freedom in the choice of the projector $\hat{P}$ and the complementary projector $1 - \hat{P}$ in the definition of the QGT in Eq.~\eqref{eq:def-Q-r-general}. Thus, the geometric tensor should not be viewed as an intrinsic property of a band or the ground state but rather as a property of the choice of projected states. These states are determined by the experimental probe, which itself imposes an effective truncation of the Hilbert space. That is, it is natural for the experimental probe to select the appropriate scale. A probe operating at a frequency range $\Delta\omega$ carries an intrinsic time uncertainty $\Delta t\approx 1/\Delta\omega$. Alternatively, the Coulomb interaction strength $U$ can impose the energy window, defining the projectors $\hat{P}$ and $1-\hat{P}$ and, therefore, which states are truncated. Unlike the Chern number, which is topologically robust and does not change when trivial bands are truncated, the quantum metric is sensitive to such truncations. Therefore, these choices, and the resulting interpretation of geometric quantities, must be critically considered when approaching a condensed matter problem.

This section is divided into three parts. The first part uses optical sum rules to establish the connection between response and geometry, and with this identify characteristic geometric length and energy scales. Building on this insight, we then review experiments capable of probing quantum geometry, with a clear identification of the associated projected subspace. Finally, focusing on a set of low-energy bands, we examine the effects of quantum geometry in many-body calculations.

\subsection*{Instantaneous response and sum rules}
The magnitude of dipole fluctuations in a real material depends on the time scale. In the long time limit, $t>\tau$, the dynamics of a single electron in a metal is diffusive with $\langle r(t) r(0) \rangle \approx Dt$ where $D$ is the Drude weight dictated by the band effective mass and carrier density. Practically, this diffusive propagation of fluctuations saturates at the mean free path, $\lambda^2 \approx D\tau$.
Note that for momentum preserving, clean metals $\lambda\to\infty$. Conversely, the very short-time dynamics $t<\hbar/V$ is governed by the local quantum chemistry of the atom. 

Quantum geometry introduces a length scale that lies in between these well-understood scales. It is comparable to the bandwidth of the optically active bands and is relevant for dynamics at times shorter than $\tau$ but larger than $\hbar/V$.
We can formalize this intuition by rewriting the Souza-Wilkens-Martin (SWM) sum rule in terms of a Fermi surface (Drude) contribution $g_{\rm FS}$, and a quantum geometric part due to sharing electrons across multiple orbitals $\ell_g^2$
\begin{equation}
   \dfrac{\hbar}{\pi n e^2} \int\limits_{0}^\infty d\omega \dfrac{{\rm Re}[\sigma_{\mu\mu}(\omega)]}{\omega} = {g_{\rm FS}\over n} + \ell_g^2 ,\label{eq:swm}
\end{equation}
where $\sigma_{\mu\mu}(\omega)$ is the longitudinal optical conductivity along one spatial direction. We have introduced a normalization with respect to charge density $n$ to make a direct comparison of this scale of dipole fluctuations to the average inter-atomic distance $a \equiv n^{-1/d}$. This gives units of length squared to the integral at every dimension. The Fermi surface contribution, $g_{\rm FS}$, is either infinite or zero, depending on whether a Fermi surface is present or absent \cite{Resta2011}.
There is no such divergence for insulators, and the largest scale of dipole fluctuations is controlled by the wavefunctions,  $\ell_g$.
Note that even in the absence of lattice interference, fluctuations exist at the smaller scale of the Bohr radius, $\ell_a\approx a_B$, which define the atomic orbitals. Capturing these requires integrating the optical conductivity up to a large energy cutoff, $V$, approximately given by the atomic level gap, to remove the contributions of core electron excitations. These high-energy transitions and fluctuations are effectively neglected in tight-binding approximations.

At intermediate frequencies between $\omega\in[\tau\inv,V]$, we can identify the geometric contribution to the localization tensor of the ground state $\ell_g$. This contribution is dominated by the band geometry without taking into account filling. 
The spectral weight at these frequencies is continuously transferred into the Drude peak and vanishes as the system is adiabatically taken to the atomic limit. This implies that it originates from quantum interference and the delocalization of orbitals across lattice sites. It can be viewed as the size of ``molecular orbitals", resonant across multiple sites in the unit cell.
Importantly, the interband optical absorption yields a finite value of $\ell_g$, both in insulators and in metals. The emergence of this scale is shown in Fig.~\ref{fig:separation-scales} with optical absorption for a free particle subjected to a honeycomb lattice potential. The finite energy contribution to optical conductivity, centered at $\E$, arises from the quantum interference between the two sublattices, implying that $\ell_g$ is comparable to the atomic distances, and the energy $\E$ can be interpreted as a resonance energy comparable to the hopping energy between sublattices. Applying Eq.~\eqref{eq:swm} to free electrons under a magnetic field of magnitude $B$, we find that $\E/\hbar=\hbar/m\ell_B^2$ is the cyclotron frequency and $\ell^2_g=\ell^2_B=\hbar/eB$ the magnetic length. However, this interpretation can be in fact generally applied to multi-orbital materials simply from the fact that bands are generally composed of admixtures of orbitals separated spatially.

Sum rules reveal that polarization fluctuations are responsible for other defining characteristics of bound electrons, such as their optical mass $1/m\approx\ev{\br\cdot\bv}$, part of orbital magnetization $M \approx \ev{\br\times\bv}$, or shot noise $\ev{\bv\cdot\bv}$. The generalized sum rules can be systematically obtained \cite{verma2024instantaneous} from different time derivatives of the unequal time dipole-dipole correlator, or time-dependent QGT, $\mathcal{Q}_{\mu\nu}(t) = \langle \hat{r}_\mu(t) (1-\hat{P}) \hat{r}_\nu\rangle $. The complementary projector $1-\hat P$ in $\Q(t)$ guarantees its gauge independence, which highly simplifies the form of response functions such as polarizability or conductivity. Taking time derivatives of $\Q(t)$ amounts to convoluting the geometric tensor with various powers of the energy differences at which the dipole transitions occur, and generalizing the SWM sum rule one finds,
\begin{align}
\int_{0^+}^\infty d\omega~{\sigma^{\rm dis}_{\mu\nu}(\omega)\over \omega^{\eta+1}} = \dfrac{\pi e^2}{\hbar} \int_{\rm BZ}\omega^\eta_{mn} \left(g^{mn}_{\mu\nu}-\dfrac{i}{2}\Omega^{mn}_{\mu\nu} \right).
\label{eq:sumRule-eta}
\end{align}
On the right-hand side, band indices, $n,m$, are summed over filled and empty states respectively, and the momentum dependence is implicit. 
On the left hand side, the dissipative conductivity contains both the longitudinal and Hall components which pick up the symmetric and  and antisymmetric parts of the geometric tensor. We introduce a low energy cutoff to remove any Fermi surface contribution, separating the finite geometric contribution of various sum rules to a possibly divergent Fermi surface part. From this we can see that if we assume that the geometric contribution to the optical conductivity is dominated by a characteristic frequency $\E/\hbar$, obtained from an weighted average of $\omega_{mn}$, we can obtain a good estimate for various instantaneous properties of insulators. 
This approximation is closely related to the Penn model, where the entire set of interband transitions is collapsed into a single effective energy scale (the ``Penn gap''), chosen to satisfy optical sum rules \cite{PennGap}.
Their optical mass may be approximated by $1/m\approx\E\ell_g^2$, shot noise by $\E^2\ell_g^2$ and the dielectric constant by $\varepsilon=1+\chi$ with $\chi\approx\ell_g^2/\E$. Larger $\ell_g$ implies larger fluctuations at all moments, and it is indeed the reason behind the large refractive index of silicon, diamond or topological insulators \cite{komissarov2024quantum}.

We should also highlight that moments with positive powers of energy differences $\hbar\omega_{mn}^\eta$, $\eta>0$ become hard to estimate without accounting for the atomic contribution at high energies. For this reason it is not possible to accurately capture orbital magnetization without involving core orbitals~\cite{Vanderbilt2018Book}, as one example. This approximation becomes exact in a system with only flat bands, such as in the Landau level problem. However, strictly speaking, all  moments of the ground state dipole fluctuations should be computed by the knowledge of all states in the Hilbert space, both filled and empty. Particularly for highly dispersive bands and broad distributions of $\sigma(\omega)$, approximating a single length scale $\ell_g$ may not be suitable. In such cases, however, we may instead keep track of these few moments as independent geometric scales, which can be used to accurately describe the response of materials beyond the single band and long-wavelength approximations, still compressing the wavefunction data to a few phenomenological numbers.

Other than estimations of geometric quantities, optical sum rules have been used to establish rigorous bounds on localization \cite{Kivelson1982, martin2020electronic, Kudinov1991, Resta1999}, superfluid stiffness \cite{Paramekanti1998, Hazra2019, Verma2021, mao2023, mao2023a}, dielectric function \cite{komissarov2024quantum, onishi2024universal}, optical mass \cite{kruchkov2023} and more recently, on the energy gap in topological insulators \cite{onishi2023quantum}.

Geometric quantities also appear in density response functions at small $q$, as the density and position operators satisfy $\hat{\rho}_q \approx 1 + i q_\mu \hat{r}_\mu$. 
This relation was recently employed to generalize sum rules to non-linear order using the density operator \cite{Bradlyn2024}.
Furthermore, it provides a pathway to extract the QGT from the static structure factor \cite{tam2024corner, Resta.Resta.200670o}, which can then be utilized to establish topological bounds \cite{onishi2024structure}.
While such sum rules and related bounds are excellent for building intuition, they are challenging to measure or verify experimentally, although some progress has been made \cite{ghosh2024probing, 2024arXiv240915583B, kang2024measurements, Kim2025}.

\subsection*{Transport coefficients}
The quantum geometry of the ground state is an essential ingredient for the response characterization of a macroscopic quantum state. In the example of diamond and rock salt, which are both insulators with comparable, large band gaps, their optical properties such as their refractive index ($n^2\approx \ell_g^2/\E$) differ significantly due to the scale $\ell_g$ being larger in covalent than ionic bonding\cite{komissarov2024quantum}.
Nevertheless, measuring quantum geometry directly remains rather complicated. 
These difficulties are intimately linked to its definition via the overlap between the projector $\hat{P}$ and the complement $1-\hat{P}$ 
(Eq.~\eqref{eq:def-Q-r-general}), which means one should probe the system adiabatically,  while simultaneously keeping track of all other energies. This somewhat paradoxical protocol is by all accounts hard to achieve but not impossible~\cite{Yi2023}. 

To give some orientation, in the following we will elucidate the main factors which enter into the integrand of a response function. However, to avoid an overly technical discussion, we will keep a loose definition of an intrinsic geometric length scale, $\ell$, (we will keep the symbol $\ell_g$ to indicate the length scale coming from the symmetric part of the QGT, the quantum metric), remembering that depending on the response, different and linearly independent geometric moments or even, combinations of moments, contribute to such $\ell$~\cite{Souza2000} which are usually tensorial objects and may differ from the definition in Eq.~\eqref{eq:swm}. We present in Table~\ref{tbl:responses} the exact forms and references regarding these contributions. 

Much progress has been made in diagnosing the most well-known part of the QGT, the Berry curvature ($\ell^2$), which can be recast as a single projection operation. 
As mentioned before, the Berry curvature leads to the intrinsic anomalous Hall effect, which describes the transverse, antisymmetric current component that arises in magnetic materials, which thus carry a nonzero momentum average of the Berry curvature. In magnetic insulators, the same mechanism gives rise to the quantum anomalous Hall effect~\cite{Haldane.Haldane.1988,Xue.Chang.2013}.
More recently, high harmonic generation and polarization-sensitive ARPES techniques were able to directly measure the momentum-resolved Berry curvature in SiO$_2$~\cite{Luu2018} and WSe$_2$~\cite{cho2021studying}. 

Another route to measuring the Berry curvature is to identify observables which couple to its momentum derivatives. 
For example, the nonlinear intrinsic anomalous Hall effect has been successfully expressed in terms of the quasiparticle lifetime times the momentum derivative of the Berry curvature ($\lambda\ell^3$)~\cite{Sodemann2015,Ma2019,Du2021}. This derivative has become known as the Berry curvature dipole,
which can retain a nonzero momentum average even in the presence of time-reversal symmetry, as long as spatial inversion symmetry is broken. The Berry curvature dipole has become a valuable probe for the Berry curvature in a number of 2d and 3d materials~\cite{Du2021}
and also offers a particularly enlightening example what ingredients can constitute the geometric length scale $\ell$. Clearly, the derivative of the Berry curvature estimates how violently the Berry curvature changes inside the Brillouin zone, setting a typical momentum space distance $\kappa$ which separates minima and maxima of the Berry curvature. In real space, this translates to a characteristic distance $\kappa^{-1} \rightarrow \ell$ below which a wavepacket remains irrotational. 
Thus, the Berry curvature dipole constitutes an anomalous component in the equation of motion that is not due to any intrinsic magnetization, but rather due to the tendency of a wavepacket to acquire a finite Hall angle from uncompensated orbital moments in the tail of the envelope function.

\begin{table*}
    \centering
    \begin{tabular}{@{}lccll@{}}
    \toprule
    Geometric quantities & Observable & Ingredients & Definition\\
    \midrule
    Berry curvature matrix elements & $\Omega_{\mu\nu}^{nm}$ & $\ell^2$ & $( r_\mu^{nm} r_\nu^{mn} - r_\nu^{nm} r_\mu^{mn} )$\\
    Quantum metric matrix elements & $g^{nm}_{\mu\nu}$ & $\ell^2$ & 
    $( r_\mu^{nm} r_\nu^{mn} + r_\nu^{nm} r_\mu^{mn} )$\\
    \midrule
    band resolved Berry curvature & $\Omega_{\mu\nu}^{n}$ & $\ell^2$ & $\sum_{m \neq n}\Omega_{\mu\nu}^{nm}$\\
    band resolved quantum metric & $g^{n}_{\mu\nu}$ & $\ell^2$ & 
    $\sum_{m \neq n}g_{\mu\nu}^{nm}$\\
    \midrule
    Berry curvature dipole & $\partial_\mu\Omega_{\nu\lambda}^{n}$& $\ell^3$ & $\sum_{m \neq n}\partial_\mu\Omega_{\nu\lambda}^{nm}$\\
    quantum metric dipole & $\partial_\mu g_{\nu\lambda}^{n}$& $\ell^3$ & $\sum_{m \neq n}\partial_\mu g_{\nu\lambda}^{nm}$\\
    Landau Zener coupling & $G_{\mu\nu}^{n}$& $\ell^2/\E$ & $\sum_{m \neq n}g_{\mu\nu}^{nm}/\omega_{nm}$\\
    Quantum connection & $Q^{nm}_{\mu\nu\lambda}$ &$\ell^3$& $ r^{nm}_\mu r^{mn}_\nu(r^{mm}_\lambda - r^{nn}_\lambda + i \partial_{\lambda} \operatorname{log} r^{mn}_\mu)$ \\
    \bottomrule
    \end{tabular}
    \caption{Examples of geometric quantities. The ``Ingredients" column keeps track of various contributions to the response: intrinsic geometric length scales $\ell$ due to position matrix elements and the intrinsic resonance energy $\E$ across bands. $m$ and $n$ label bands and $\hbar\omega_{mn}$ labels the energy difference between bands.
    }
    \label{tbl:responses}
\end{table*}

However, no direct measurements of the momentum-resolved quantum metric ($\ell^2$) or its momentum average have been reported so far. There have been suggestions that the metric might couple directly to the dc-conductivity~\cite{Mitscherling2022,Kruchkov2022,Huhtinen2023} in certain fine-tuned settings, but no linear response observable is known which is generically determined solely by the quantum metric.
It has also been suggested that the quantum metric dipole ($\ell^3$) appears in the nonreciprocal directional dichroism for materials breaking time-reversal symmetry, a nonlocal conductivity component which originates from a coupling to the electric field gradient~\cite{Lapa2019,Gao2019}.

\begin{table*}
    \centering
    \begin{tabular}{@{}lccll@{}}
    \toprule

    \strut Phenomenon&   Observable & Ingredients & Definition & Refs. \\
    \midrule
    Anomalous Hall & $\sigma_{\mu\nu}$ & $\ell^2$& 
    $\frac{e^2}{\hbar} \int_{\rm BZ}\sum_n  f_n \Omega^n_{\mu\nu}$ & \cite{Haldane.Haldane.1988,Xue.Chang.2013,Xiao2010,Nagaosa2010}\\
    Nonlinear anom. Hall& $\sigma^{(2)}_{\mu\mu\nu}$ & $\lambda\ell^3$& 
    $\tau\frac{e^3}{\hbar^2} \int_{\rm BZ}\sum_n  f_n \partial_\mu \Omega^n_{\mu\nu}$ & \cite{Sodemann2015,Ma2019,Du2021}\\
    \midrule
    Non-reciprocal dichroism& $\gamma_{\mu\nu\lambda}$ & $\ell^3$& 
    $-e^2 \int_{\rm BZ}\sum_n  f_n \partial_\nu
    g^n_{\mu\lambda}$ & \cite{Lapa2019,Gao2019}\\
    Static susceptibility& $\chi_{\mu\nu}$ & $\ell^2/\E$& 
    $\frac{e^2}{2\hbar} \int_{\rm BZ}\sum_n  f_n G^n_{\mu\nu} $ &\cite{komissarov2024quantum}\\
    Non-reciprocal conductivity& $\sigma^{(2)}_{\mu\mu\nu}$& $\ell^3/\E$& 
    $ \frac{e^3}{\hbar^2} \int_{\rm BZ}\sum_n  f_n \left( \partial_\mu G^n_{\mu\nu}-2 \partial_\nu G^n_{\mu\mu}    \right)$ &\cite{wang2023quantum,gao2023quantum,Gao2014,Liu2021,Kaplan2023,Kaplan2024}\\
    \midrule
    Optical transition rate& $\Gamma_{\mu\nu}(\omega)$& $\ell^2/\omega$ & 
    $ \pi\int_{\rm BZ}
    \sum'_{nm}
    g^{mn}_{\mu\nu} \delta (\omega - \omega_{mn})$ &\cite{Chen2022,deSousa2023a}\\
    Current injection (linear pol.)& $\sigma^{(2)}_{\mu\mu\lambda}(\omega)$& $\lambda\V\ell^2$ & 
    $ \tau \frac{2 \pi e^3}{\hbar^2}  \int_{\rm BZ}
    \sum'_{nm}
    \partial_\lambda\omega_{mn} g_{\mu\nu}^{nm} \delta(\omega - \omega_{mn})$  & \cite{Holder2020,Ma2021,Nagaosa2022,Jiang2025}\\
    Current injection (circular pol.)& $\sigma^{(2)}_{\mu\nu\lambda}(\omega)$& $\lambda\V\ell^2$ & 
    $\tau \frac{2 \pi e^3}{\hbar^2}  \int_{\rm BZ}
    \sum'_{nm}
    \partial_\lambda\omega_{mn} \Omega_{\mu\nu}^{nm} \delta(\omega - \omega_{mn})$  & \cite{deJuan2017,Ma2021,Nagaosa2022,Jiang2025}\\
    Shift current & $\sigma^{(2)}_{\mu\nu\lambda}(\omega)$ & $\ell^3$& 
    $ \frac{i\pi e^3}{-2\hbar^2} \int_{\rm BZ}
    \sum'_{nm}
    (Q^{nm}_{\mu\nu\lambda}-Q^{nm *}_{\nu\mu\lambda})
    \delta (\omega - \omega_{mn})
    $ & \cite{Ahn2022,Avdoshkin2024}\\
    \midrule
    Spectral weight${}^\dag$ & $n/m_\lambda$ & $\ell^2\E$ & $(1/\hbar)\int_{\rm BZ}\sum^*_{nm} 
    \omega_{mn}g^{nm}_{\lambda\lambda}$ & \cite{resta2017geometrical}\\
    Orbital magnetization${}^\dag$ & $M_\lambda$ & $\ell^2\E$ & 
    $(e/c)\int_{\rm BZ}\sum^*_{nm}  \omega_{mn}\Omega^{nm}_{\mu\nu}\epsilon_{\mu\nu \lambda}$ & \cite{Souza2008}\\
    Many-body metric${}^\dag$ & $g_\lambda$ & $\ell^2$ & 
    $(\hbar/\pi e^2)\int_{\rm BZ}\sum^*_{nm}   g^{nm}_{\lambda\lambda}$ & \cite{Souza2000} \\
    Chern number & $\mathcal{C}_\lambda$ & $\ell^2$ & 
    $\epsilon_{\mu\nu \lambda} \int_{\rm BZ}\sum^*_{nm}   \Omega^{nm}_{\mu\nu}$ & \cite{TKNN1982}\\
    \bottomrule

    \end{tabular}
    \caption{Examples of mixed transport coefficients and some optical responses. Definitions:
    Occupation function $f_n$, and the double sum over occupied and empty states is defined as $\sum^*_{nm}=\sum_{n,m \neq n}f_n(1-f_m)$, while $\sum'_{nm}=\sum_{n,m \neq n}(f_n-f_m)$. The symbol $\epsilon_{\mu\nu\lambda}$ is the Levi-Civita totally antisymmetric tensor and $\delta_{mn}$ the Kronecker delta. 
    The ``Ingredients" column now includes velocities $\V$ in addition of the intrinsic geometric length scale $\ell$ and $\E$.
    We also highlight external factors such as as an external frequency $\omega$ dictating resonant responses or a momentum relaxation mean free path $\lambda$ due to disorder or interactions. ${}^\dag$Only the interband, geometric contributions.}
    \label{tab:tab3}
\end{table*}

Even though the quantum metric of the ground state seems rather elusive, the corresponding dipole matrix elements appear almost ubiquitously in response functions in the form of wavefunction overlaps.
These matrix elements may have origins in quantum geometry or simply due to onsite atomic transitions, and are generally weighted by velocities or energies. The resulting observables thus depend on both quantum geometry and band energies. In these cases we must introduce a resonance energy scale $\E$ defined by the band dispersion of the material, as it was done in the previous section.
To give one example, in a multi-orbital system, the nonlocal conductivity acquires a complicated dependence on several dispersive and geometric features~\cite{Kozii2021}.

Many cases of these mixed responses have been reported, but particularly successful have been examples where the interband quantum metric is normalized by the band gaps ($\ell^2/\E$). Such a term accounts for the unavoidable mixing of bands in the presence of an adiabatic perturbation. This is highly analogous to the Landau-Zener effect~\cite{Weinberg2017} and is thus most appropriately viewed as the excitation of virtual interband transitions~\cite{Holder2021a}.
Landau Zener mixing can be identified in the low frequency linear capacitance of insulators, which has been employed to illuminate the role of the QGT for dielectric properties~\cite{komissarov2024quantum}.

A similar effect can be observed in the nonlinear dc-conductivity of time-reversal breaking band structures~\cite{Morimoto2023,Jiang2025}, leading to  
a novel nonlinear Hall effect and non-reciprocal longitudinal conductivity~\cite{wang2023quantum,gao2023quantum}. Both contributions originate from the dipole of the Landau-Zener mixing ($\ell^3/\E$)~\cite{Gao2014,Liu2021,Kaplan2023,Kaplan2024}.
Recently, the third order conductivity has been derived for the study of altermagnets, which yields structurally analogous mixed response functions with Berry curvature and Landau-Zener terms as building blocks, albeit in more complicated combinations ($\ell^4/\E$)~\cite{Liu2022a,Fang2024}.

Finally, one can search for responses which contain the band-resolved QGT in a given response function. This is often possible in resonant optical responses, defined by an external photon frequency $\omega$ of the incident light.
Notably, the interband matrix elements from optical transition rates (Fermi's golden rule) yield the QGT, but evaluated with respect to two energies which are separated by the photon energy of the incident light ($\ell^2/\omega$)~\cite{Chen2022,deSousa2023a}.
Several examples of resonant mixing have also been reported for the nonlinear optical conductivity~\cite{Ma2021,Nagaosa2022,Jiang2025}. Namely, there is a quantized circular photogalvanic effect (cf. circular polarized current injection in Table~\ref{tab:tab3}) in two-band systems originating from the Berry charges located at Weyl points~\cite{deJuan2017}, whose size $\lambda\ell^2$ is the result of a convenient cancellation of the Fermi velocities with the integration measure. 
More generally, shift ($\ell^3$) and injection currents ($\lambda\V\ell^2$) have been connected
to matrix elements of the QGT~\cite{Holder2020} and to geometric objects which go beyond the QGT, which has been termed Riemannian geometry
~\cite{Ahn2020,Ahn2022,Bouhon2023,Jankowski2024} or multi-state geometry~\cite{Avdoshkin2023,Avdoshkin2024}.
Experimentally, bulk photovoltaic responses have been prominently observed in semimetals like CoSi~\cite{Ni2021} and TaAs~\cite{Ma2017,Osterhoudt2019}.

As an illustration, let us analyze the the longitudinal injection current (Table~\ref{tab:tab3}). It features three  ingredients, (i) the difference of the band velocities between valence and conduction band, (ii) a delta function which enforces resonance and (iii) the matrix elements of the interband quantum metric. It is therefore possible to combine (i) and (ii) in order to write the injection current as the integral over the quantum metric over the space of optically allowed transitions $\vec{S}_k$,
\begin{align}
\sigma^{(2)}_{\mu\mu\nu} & 
= - \tau \frac{2 \pi e^3}{\hbar^2} \hat{\nu} \cdot \int_{\omega_{mn}=\omega}\!\!\! d\vec{S}_k g^{\mu\mu}.
\end{align}
As discussed before, this integral measures properties of the interband quantum geometry, not the ground state metric itself. 

The overall scaling of the response, $\sigma^{(2)}_{\mu\mu\nu}\approx\lambda\V\ell^2$, can be understood from three physical ingredients. First, the response is proportional to the mean free path $\lambda$ because the injected carriers must scatter to form a steady DC current. Second, the strength of the resonant optical transition, driven by a photon of frequency $\omega$, is governed by the quantum metric, which provides an area scale $\ell^2$. Third, a net current requires an asymmetry in the band structure, provided by the velocity difference between bands, which contributes the final intrinsic scale $\V$.

Injection currents require a mismatch in band velocities between time-reversed processes; an intuitive interpretation can therefore be given in terms of the two triangle diagrams that constitute this response, which are related by time reversal~\cite{Holder2020,Jiang2025}. In the absence of time reversal, the semiclassical quasiparticle motion in the clockwise and anti-clockwise nonlinear excitation does not occupy identical real-space areas because they involve different velocities. The effect is thus proportional to the transition area $\ell^2$ times the intrinsic velocity mismatch, which depends on the external frequency $\V$.

Despite the tremendous progress in the last few years in characterizing geometric features in observables, it is important to stay parsimonious in applying this label.
Matrix overlaps are ubiquitous in multi-orbital systems, and not every occurrence of a matrix element automatically signals that quantum geometry (dipole fluctuations from orbital frustration and delocalization) is an important ingredient towards understanding a particular response. We should be aware of the fact that the momentum structure of the matrix overlaps leads in many cases only to quantitative modifications, without changing the overall character of the response.  

Nonetheless, the examples listed above show that it is possible to distill and isolate the role of quantum geometry, and we are confident that many more examples like these will be found. 
Another important aspect is that almost all of the response functions discussed above can additionally receive large contributions from extrinsic origin ($\lambda$), like impurity scattering, phonons and boundary effects~\cite{sturman2021photovoltaic,Du2021,Atencia2022}. It is an ongoing challenge to disentangle the latter from the intrinsic phenomena ($\ell$) which originate from the bulk wavefunctions, and this is only possible on a case-by-case basis.

\subsection*{Correlated phases}

The SWM sum rule and the localization tensor have been known for over three decades. 
More recently, the surge of interest in quantum geometry has been driven by the discovery of the magic angle in twisted bilayer graphene \cite{cao2018, cao2018a, Yankowitz2019, Lu2019} and the rapid development of moiré materials \cite{Andrei2021}. 
These systems exhibit flat bands due to destructive interference over large moiré unit cells $a_m$, for which the geometric length scale is typically $\ell_g \approx a_m \gg a$. 
While moiré materials clearly host significant geometric effects, much of the existing literature pursues questions that differ conceptually from the theme of this review. 
Our focus has been on ground-state dipole fluctuations, whereas in moiré systems the interest lies in how such fluctuations reshape interactions and thereby stabilize correlated phases.

Nevertheless, viewing the problem through the lens of a geometric length scale provides useful intuition. 
Dipole fluctuations are tied to extended real-space charge distributions, which can cause even short-range interactions to acquire spatial structure at the geometric scale. 
In practice, these effects enter through projected interactions,
\begin{equation}
    \bar{H}_{\rm int} = \sum_q V_q \bar{\rho}_q \bar{\rho}_{-q}, 
    \qquad 
    \bar{\rho}_q = \sum_k \langle u_k | u_{k+q} \rangle \, c^{\dagger}_{k+q} c_k ,
\end{equation}
where the form factors $\langle u_k | u_{k+q}\rangle$ encode the quantum geometry.

A central conceptual point is that projected interactions depend on the form-factor structure at all $q$.  
This differs fundamentally from the small-$q$ overlaps that define the projected position operator and the quantum geometric tensor \cite{Antebi2024}. 
Because interactions probe higher moments of the dipole operator, isolating contributions that depend only on the small-$q$ geometric tensor is generally difficult.  
By dimensional analysis, any observable depending solely on the quantum metric must introduce an additional length scale to compensate for the extra factor of $q$ relating projected density to projected position.

A remarkable example where small-$q$ geometry becomes decisive is the stabilization of lattice fractional Chern insulators.  
The form factors provide a direct route to understanding this mechanism \cite{ParameswaranRoySondhi2013}.  
A key result is the trace condition \cite{Roy2014}, which fixes the relation between the metric and Berry curvature and constrains fluctuations of the vortex operator $z=x+iy$. 
Satisfying this constraint is sufficient for obtaining exact fractional Chern insulator ground states for short-range pseudopotentials, in analogy with Laughlin’s construction \cite{Ledwith2023, Wang2021ExactFlatband}.  
Related notions, such as ideal quantum geometry \cite{Ledwith2021StrongCouplingGraphene}, Kähler geometry \cite{MeraOzawa2021kahler}, and vortexability \cite{Ledwith2023}, are alternative formulations that have gained prominence because in the chiral limit of twisted bilayer graphene \cite{Tarnopolsky2019}, the wavefunctions share similarities with the lowest Landau level \cite{Ledwith2020, Wang2021ExactFlatband}.  
The broader role of geometry in fractional Chern insulators continues to develop, especially following recent observations of fractional states in moiré systems \cite{Xie2021, Cai2023, Zeng2023}.

Another setting where geometric effects manifest cleanly is flat-band superconductivity. 
Based solely on dimensions,
the $f$-sum rule of Eq.~\eqref{eq:sumRule-eta} (with $\eta = 1$) is a good candidate for quantum geometric contributions.
Although the sum rule carries energetic prefactors unrelated to the quantum metric, interactions can replace these with an interaction energy scale, enabling a finite low-energy sum rule even in a perfectly flat band.  
Because the superfluid stiffness is bounded above by this sum rule \cite{Zhang.Scalapino.1993}, a nonzero stiffness becomes possible in flat bands.

At a microscopic level, however, the superfluid stiffness generally depends on the full form-factor structure of the projected interaction, not solely on small-$q$ dipole fluctuations.
Although this makes the problem sensitive to model-specific details, a particularly clear illustration was provided in Ref.\cite{PeottaTorma2015}, which showed that in flat-band lattice models, even a perfectly flat band can support a finite superfluid stiffness when attractive interactions and quantum geometry are both present.
In their calculation for a multi-orbital attractive Hubbard model with the uniform pairing condition \cite{Herzog2022a} and within single-mode approximation, the stiffness scales as $D_s \propto n U \ell^2$, where $n$ is the charge density and $\ell^2$ is a geometric length scale given by the trace of the quantum metric of the flat band \cite{Julku2016, torma2022}. 
Using projection based on Wannier functions, it was shown that this effect is a consequence of projected interactions containing a pair-hopping process with effective charge $2e$, whose strength is highly sensitive to the underlying quantum geometry~\cite{Tovmasyan2016EffectiveModels}.
Subsequent works reformulated this phenomenon in terms of position operators, emphasizing that unlike the full density operator, the projected density couples to external gauge fields and hence leads to a non-vanishing low-energy optical response~\cite{Verma2021, mao2023, mao2023a}. 

The precise scaling between the superfluid stiffness and the geometric length scale can be refined using the generalized random phase approximation~\cite{Tam2024}, which preserves the same functional form but includes the minimal quantum metric \cite{Huhtinen2022,Lamponen2025,Buthenhoff2025}. 
Physically, the stiffness sets the effective Cooper-pair mass, so the relation $1/m_{\rm eff} \approx U \ell^2$, can be viewed as the geometric length scale renormalizing the pair mass, as confirmed from a calculation of the two-body bound state with separable interaction–potentials \cite{Torma2018}.
Quantitative estimates of this renormalization have been obtained in twisted bilayer graphene \cite{hu2019, XiePRL2020, JulkuPRB2020}. 
Beyond stiffness, several other superconducting properties exhibit quantum geometric contributions, including the coherence length \cite{Chen2024}, the Josephson current \cite{Law.Li.2025}, and the localization of Majorana zero modes \cite{Law.Guo.2024}.

More broadly, the interplay of quantum geometry and correlations is not restricted to fractional Chern insulators or superconducting ground states. 
Similar effects appear as well in systems where the ground state supports excitons \cite{Xu2022, rossi2021quantum, Fertig1989, Verma2024} or magnetic order, with geometry contributing to flat-band ferromagnetism \cite{Tasaki1998} and spin stiffness \cite{WuPRBspinstiffness}.
Finally, we emphasize that a correlated ground state is not a prerequisite: even in weakly interacting bands the interband Coulomb interaction, together with the interband current operator, can renormalize the carrier effective mass \cite{Abedinpour2011, Li2017, Antebi2024} and optical conductivity \cite{Tse2009, CarmichaelClaassen2025}, though this is yet to be observed experimentally.

We note that a large geometric length scale does not always lead to exotic correlated phases or unusual transport. We have in this review focused on $\ell_g$ appearing from lattice interference at the unit cell scale, but polarization fluctuations may appear through other sources of interference. Metals with impurities provide a simple illustration\cite{komissarov2025superdielectrics}. 
Impurities introduce an effective random lattice that reduces the quantum metric from infinity to a large finite value given by the mean free path, $\lambda\gg a$, yet this change leaves the metallic state essentially unchanged and often well described withing semiclassical approximations and Fermi liquid theory. 
Higher Landau levels offer another counter-example in a similar vein. The quantum metric grows with the Landau level index $N$, but at large $N$, the system enters a semiclassical regime where bubble and stripe phases appear rather than fractional Chern insulators \cite{ChalkerMoessner1996,Fogler2002StripesBubbles}.
Therefore we should be careful to evaluate in every case whether geometry is reshaping phenomena in a meaningful way.

\section*{Outlook}

Quantum interference and orbital mixing across distinct lattice sites introduce electron dynamics at scales comparable to (and sometimes larger than) the lattice constant, fundamentally modifying the physical properties and phases of quantum materials. Quantum geometry provides a systematic framework for capturing the phenomenology arising from dipole fluctuations. As we discussed above, these effects can be understood by the length and time scales which are imposed not by the band dispersion, but by the momentum dependence of the electron wavefunctions.
Therefore, we advocate revisiting minimal models of quantum materials from this fresh perspective. 
We should start by estimating the magnitude of geometric effects in materials with exotic physical properties. As described here, these scales can be quickly estimated using the optical conductivity (Fig.\ref{fig:separation-scales})  both in \emph{ab-initio} computations and direct measurement, and complemented by various other observables organized in Table~\ref{tab:tab3}. 

That said, many fundamental questions remain open, and the full implications of quantum geometry in solid state physics are in its infancy. Much of the focus on quantum geometric effects have been given to insulators and flat bands \cite{torma2022}, when in fact these are universal effects that dictate the impact of having multiple orbitals. Metals may offer the next natural setting to explore quantum geometric effects, where they could play a role in driving instabilities toward a particular correlated phase. While the ground-state metric diverges in a metal, recent works particularly in kagome metals \cite{DiSante2025KagomeMetals}, suggests that nontrivial wavefunction geometry along the Fermi surface can strongly influence susceptibilities to ordered phases. This motivates a re-examination of how quantum geometry might renormalize electron dynamics at the Fermi surface, although its precise connection to traditional frameworks like Landau's Fermi liquid theory remains an open question.

Here we list several key questions and directions we hope to see developments in the future:

\subsection*{Better models} 
Quantum geometry quantifies the effects of Hilbert space truncation in effective theories, providing a systematic framework for including corrections that arise from orbital mixing. 
This can lead to more accurate effective models that capture the relevant scales without requiring the full atomic basis. If we are presented with materials with many orbitals and, at first glance, they cannot be disentangled, how do we answer the question: which orbitals matter? 

\subsection*{Materials discovery.}  

Understanding the quantum geometry of materials and how it emerges by specific lattice motifs can guide the search for materials with enhanced responses or novel quantum phases. How can we utilize the vast computational capabilities of material science to espouse such geometric insights? Can we guess which materials will have which order based not only on fermiology but also on geometry?

\subsection*{Low temperature phenomena.}
A key open challenge is understanding how interactions combined with significant dipole fluctuations affect the dynamics of electrons around the Fermi surface in metals. Can these effects be quantified through electronic transport at the Fermi surface, such as in its quantum oscillations? 
It is now recognized that even weak interactions $ U \ll \E $ can transfer spectral weight from high frequencies ($ \hbar\omega \gg \E $) to low frequencies, thereby enhancing charge transport, both in metals~\cite{XuYang2025} and in correlated condensates~\cite{torma2022, rossi2021quantum}. However, whether this redistribution reflects a universal geometric mechanism remains an open question. In other words, can quantum geometry universally ``speed up'' charge carriers in quantum matter? If so, what experiments could isolate carrier dynamics with and without geometric contributions?

\subsection*{Unifying Landau level and band phenomena.}
Landau levels have extraordinary properties and are known to stabilize many exotic quantum states such as fractional topological order or exciton condensates \cite{Eisenstein2014}. However, it comes at the expense of high magnetic fields. The geometric framework introduces a natural frequency $\E/\hbar$ and an intrinsic geometric length $\ell$ that mimic the cyclotron frequency and magnetic lengths, however these emerge from lattice interference and not from the external Lorentz force. With the recent advent of moiré heterostructures, many quantum Hall phenomena have been mimicked in the absence of a magnetic field, opening a myriad of possibilities for new quantum devices. However, moiré systems come with their own problem of scalability. The proof of concept has been done - can we now find scalable materials with exotic orders which can be incorporated in quantum devices?

Let us conclude with the question: is it necessary to include quantum geometry in the standard toolbox? As physicists, we often look for the simplest description that abstracts away irrelevant degrees of freedom. Quantum geometry quantifies this process by allowing us to define energy and length scales with origins in orbital mixing.
In modeling any condensed matter system, we strive to find a middle ground between complexity and simplicity, a compromise that dictates the level of description we adopt. Indeed, there is no ambiguity in defining the many-body Hamiltonian at the UV scale - the kinetic energy depends on the momentum operator, and interactions depend on the density operator.
It is only when the theory is projected onto a subset of states that quantum geometric features emerge, often leading to low-energy theories that defy naive semiclassical intuition, as seen in flat-band superconductors and fractionalized quantum Hall states \cite{Haldane.Haldane.2011}.
Quantum geometry is thus a natural and essential byproduct of our reliance on effective theories. Tying back to the notion of dipole fluctuations, this perspective can be succinctly summarized as follows:
If the ground state of a material has strong orbital mixing, it cannot be described simultaneously on the basis of a single local orbital and in flat space. We can either choose a multi-band picture in flat space or a single-band picture in a curved geometry.

\section*{ Box 1: A case study of transition metal dichalcogenides}

Transition metal dichalcogenide semiconductors exemplify many of the central themes of this review. 
They share the chemical formula $\rm MX_2$, where $\rm M$ is a transition metal ($\rm Mo$, $\rm W$) and $\rm X$ is a chalcogen ($\rm S$, $\rm Se$, $\rm Te$). 
Each monolayer consists of a hexagonal lattice of transition metal atoms sandwiched between two chalcogen planes, forming a quasi-2D crystal with strong in-plane covalent bonds and weak interlayer van der Waals interactions. 
For broad overviews of their chemistry, band structure, and optical properties, see Refs.~\cite{Wang2018_RMP, Manzeli2017_NatRevMat}. Like graphene, whose physical properties are well understood to stem from the geometry of its Dirac points, these materials show a remarkable zoo of exotic correlated phases and responses which can be attributed to the nontrivial structure of their wavefunctions.

Unlike semiconductors such as {\rm GaAs} where the effective mass approximation applies well, minimal tight-binding models of transition metal dichalcogenides cannot be reduced to a single-orbital. 
Rather, the conduction and valence bands derive from the bonding of three hybrid orbitals across neighboring unit cells, localized mostly in the empty space between metal and chalcogen \cite{Holbrook2024RealSpaceImaging}. This originates from an effective lattice frustration and results in a substantial flattening of the valence band. In these semiconductors, dipole fluctuations are large and extended beyond one unit cell. This topological obstruction \cite{Yu_deJuan_2025_QuantumGeometryNbSe2I} produces very large valley-contrasting Berry curvature \cite{Xiao2012_PRL_Berry, Xu2014_NatPhys_valley}, spin–orbit splitting \cite{Zhu2011_PRB_SOC, Zeng2012_NatNano_valleyPol}, and a nontrivial band geometry\cite{Kormanyos2015_kp}.

Because of this intrinsic geometry,
both Landau levels \cite{Xuan2020,ding2025cyclic} and excitons \cite{Chernikov2014} have long been observed to deviate from the effective mass approximation, with some of these deviations attributed to quantum geometric contributions \cite{Zhou2015, Srivastava_2015}. In the optical domain, this same quantum geometry underlies the giant nonlinear responses \cite{Wen2019}, including unusually strong second-harmonic generation \cite{Yin2014TMD}. All these effects can, retrospectively, be attributed to abnormally large dipole fluctuations, due to their multi-orbital character.

Twisting or stacking these materials into moiré heterostructures amplifies these effects by producing flat bands in which geometric contributions dominate the interactions.  
Experimentally, such systems host a remarkably wide range of correlated phenomena, including superconductivity \cite{Guo2024}, fractional quantum Hall phases \cite{Cai2023, Zeng2023, Park2023, Xu2023}, quantum anomalous Hall effects \cite{Li2021a}, generalized Wigner crystals \cite{Regan2020, Li2021}, and other symmetry-broken insulating states \cite{Wang2020, Tang2020, Xu2022a, Anderson2023, Xu2020}. On the theoretical side, the role of quantum geometry in these moiré bands has been elucidated through studies of form factors \cite{Dong2023CFLtMoTe2, MoralesDuran2023MagicLine, MoralesDuran2024PRL, Shi2024ACBands, Abouelkomsan2023QMetricMoires, verma2025local} and chiral limits \cite{Crepel2024ChiralLimitTMD}.  
For a broader overview of this rapidly growing field we refer readers to Ref.~\cite{Yu2024QuantumGeometry}.

\subsection*{Acknowledgments}
R.Q. and N.V. acknowledge support from the U.S. Department of Energy through the Center for Programmable Quantum Materials (DE-SC0019443). R.Q. also acknowledges support from the National Science Foundation CAREER program (DMR-2340394) and the Alfred P. Sloan Foundation (FG-2025-24714). 
PJWM acknowledges support by the European Research Council (ERC) under the European Union’s Horizon research and innovation programme (XBEND, grant agreement no. 101080740).
T.H. acknowledges financial support by the European Research Council (ERC) under grant QuantumCUSP
(Grant Agreement No. 101077020). 

\subsection*{Author contributions}
All authors contributed to the conceptualization, writing, and editing of the manuscript.

\bibliography{qgeometry_review}

\end{document}